\documentclass[10pt]{iopart}
\usepackage{graphicx,epsfig,iopams}
\begin{document}

\title{Testing the No-Hair Theorem with Sgr A$^*$}

\vspace{0.8cm}

\author{Tim Johannsen}
\address{Physics Department, University of Arizona, 1118 E. 4th St., Tucson, AZ, 85721, USA}
\eads{\mailto{timj@physics.arizona.edu}}

\begin{abstract}

The no-hair theorem characterizes the fundamental nature of black holes in general relativity. This theorem can be tested observationally by measuring the mass and spin of a black hole as well as its quadrupole moment, which may deviate from the expected Kerr value. Sgr~A$^*$, the supermassive black hole at the center of the Milky Way, is a prime candidate for such tests thanks to its large angular size, high brightness, and rich population of nearby stars. In this review I discuss a new theoretical framework for a test of the no-hair theorem that is ideal for imaging observations of Sgr~A$^*$ with very-long baseline interferometry (VLBI). The approach is formulated in terms of a Kerr-like spacetime that depends on a free parameter and is regular everywhere outside of the event horizon. Together with the results from astrometric and timing observations, VLBI imaging of Sgr~A$^*$ may lead to a secure test of the no-hair theorem.

\end{abstract}


\section{Introduction}

According to the no-hair theorem, black holes are uniquely characterized by their masses and spins and are described by the Kerr metric \cite{NHT}. Mass and spin are the first two multipole moments of the Kerr spacetime, and all higher order moments can be expressed in terms of these two \cite{Multipoles}. The no-hair theorem, then, naturally leads to the expectation that all astrophysical black holes are Kerr black holes. To date, however, a definite proof for the existence of such black holes is still lacking despite a wealth of observational evidence (see discussion in, e.g., \cite{Psaltis06}).

Tests of the no-hair theorem have been suggested using observations in either the gravitational-wave [4-9]
 or the electromagnetic spectrum [10-16]
. Both approaches are based on parametric frameworks that contain one or more free parameters in addition to mass and spin which measure potential deviations from the Kerr metric [6-8,17-19].
 If no deviation is detected, then the compact object is indeed a Kerr black hole. However, since such deviations can have a significant impact on the observed signals, the no-hair theorem may be tested in a twofold manner: If a deviation is measured to be nonzero and if general relativity is assumed, the object cannot be a black hole \cite{CH04,Hughes06}. Alternatively, if the object is otherwise known to possess an event horizon, it is a black hole, but different from a Kerr black hole. In the latter case, the no-hair theorem would be falsified \cite{PaperI}.

Sgr~A$^*$, the supermassive black hole at the center of the Milky Way, is a prime target for testing strong-field gravity and the no-hair theorem with electromagnetic observations (see \cite{Review} for a review). Monitoring the orbits of stars around this compact object for more than a decade has lead to precise mass and distance measurements making Sgr~A$^*$ the black hole with the largest angular size in the sky \cite{Sgr}. In addition, very-long baseline interferometric observations have resolved Sgr~A$^*$ on event horizon scales \cite{Doeleman08}. On the theoretical side, there have been significant advances recently in the development of a framework within which the search for violations of the no-hair theorem can be carried out.

In this article, I review this framework as well as the prospects for an observational test of the no-hair theorem with Sgr~A$^*$.

\section{An Ideal Framework for Testing the No-Hair Theorem}

Spacetimes of rotating stellar objects in general relativity have been studied for several decades. Due to the nonlinearity of Einstein field equations, the construction of such metrics is plagued with sometimes incredible technical challenges. Following the discovery of the Schwarzschild \cite{Schwarzschild16} and Kerr metrics \cite{Kerr63} in 1916 and 1963, respectively, Hartle and Thorne \cite{HTmetric} constructed a metric for slowly rotating neutron stars that is appropriate up to the quadrupole order. Tomimatsu and Sato \cite{TSmetrics} found a discrete family of spacetimes in 1972 that contain the Kerr metric as a special case. After a full decade of research, Manko and Novikov \cite{MN92} found two classes of metrics in 1992 that are characterized by an arbitrary set of multipole moments.

Many exact solutions of the Einstein field equations are now known \cite{Stephani03}. Of particular interest is the subclass of stationary, axisymmetric, vacuum (SAV) solutions of the Einstein equations, and especially those metrics within this class that are also asymptotically flat. Once an explicit SAV has been found, all SAVs can in principle be generated by a series of HKX-transformations (\cite{HKX} and references therein), which form an infinite-dimensional Lie group \cite{GerochGroup}. Each SAV is fully and uniquely specified by a set of scalar multipole moments \cite{Beig} and can also be generated from a given set of multipole moments \cite{Sibgatullin}. These solutions, however, are generally very complicated and often unphysical. For some astrophysical applications, such as the study of neutron stars, it is oftentimes more convenient to resort to a numerical solution of the field equations \cite{numerical}.

To date, there exist seven different approaches that model parametric deviations from the Kerr metric. Ryan \cite{Ryan95} studied the motion of test particles in the equatorial plane of compact objects with a general expansion in Geroch-Hansen multipoles. Collins \& Hughes \cite{CH04}, Vigeland \& Hughes \cite{VH10}, and Vigeland et al. \cite{Vigeland11} constructed Schwarzschild and Kerr metrics with perturbations in the form of Weyl sector bumps. Glampedakis \& Babak \cite{GB06} designed a metric starting from the Hartle-Thorne metric \cite{HTmetric} that deviates from the Kerr metric by an independent quadrupole moment. Gair et al. \cite{Gair08} applied a similar technique to the Manko-Novikov metric \cite{MN92} affecting the quadrupole as well as higher order moments. Sopuerta \& Yunes \cite{SY09} found a metric for a slowly rotating black hole that violates parity. Vigeland et al. \cite{Vigeland11} designed parametric deviations from the Kerr metric that possess four integrals of the motion and, hence, allow for the full separability of the Hamilton-Jacobi equations. Finally, Johannsen \& Psaltis constructed a metric of a rapidly rotating Kerr-like black hole \cite{JP11}. Other metrics of static black holes in alternative theories of gravity have also been found (e.g., \cite{staticblackholes}).

Due to the no-hair theorem, the Kerr metric is the only asymptotically flat SAV in general relativity with an event horizon but no closed timelike loops \cite{NHT}. Consequently, any parametric deviation within general relativity has to violate at least one of these prerequisites and introduces either singularities or regions with closed timelike loops outside of the event horizon, which usually occur very near to the central object at radii $r\lesssim2M$ \cite{Johannsen11}. Otherwise, these metrics would render the no-hair theorem false. The relevance of this kind of pathologies depends on the astrophysical application. They play no role for tests of the no-hair theorem that only involve the orbits of objects at large distances from the horizon, as is the case for extreme mass-ratio inspirals or the motion of stars or pulsars around a black hole. They are, however, critical for the study of accretion flows around black holes \cite{JP11}, because the electromagnetic radiation originates predominantly from the immediate vicinity of the event horizon.

For this reason, the emission from accretion flows around black holes is most interesting for strong-field tests of the no-hair theorem with observations across the electromagnetic spectrum ranging from X-ray observations of quasi-periodic variability, fluorescent iron lines, or continuum disk spectra \cite{PaperI,PaperIII} to sub-mm imaging of supermassive black holes with VLBI \cite{PaperI,PaperII}. All of these observation techniques critically depend on the location of either the circular photon orbit or the innermost stable circular orbit (ISCO), because these orbits dominate the characteristics of the received signals.

These strong-field tests of the no-hair theorem require a very careful modeling of the inner region of the spacetime of black holes. Due to the pathologies of previously known parametric deviations, it has been necessary to impose an artificial cutoff at some radius outside of the event horizon that encloses all of the above pathologies and, thereby, shields them from the observer. Therefore, the application of parametric frameworks to such tests of the no-hair theorem in the electromagnetic spectrum has, so far, been limited to only slowly to moderately spinning black holes, for which the circular photon orbit and ISCO are still located outside of the cutoff radius \cite{JP11,Johannsen11}.

Recently \cite{JP11}, we constructed a black hole metric that is regular everywhere outside of the event horizon for all values of the spin within the allowable range and that depends on a set of free parameters in addition to mass and spin. In the case when all parameters vanish, our metric reduces smoothly to the Kerr metric. Our metric is a vacuum solution of a more general set of field equations, but otherwise fulfills all of the prerequisites of the no-hair theorem and, therefore, preserves these essential properties even if the deviation parameters from the Kerr metric are nonzero. At present, our metric constitutes the only known black hole spacetime of this kind and serves as an ideal framework for the study of the signatures of a possible violation of the no-hair theorem from astrophysical phenomena near the event horizon of a black hole and, in particular, of Sgr~A$^*$.

\begin{figure}[t]
\begin{center}
\psfig{figure=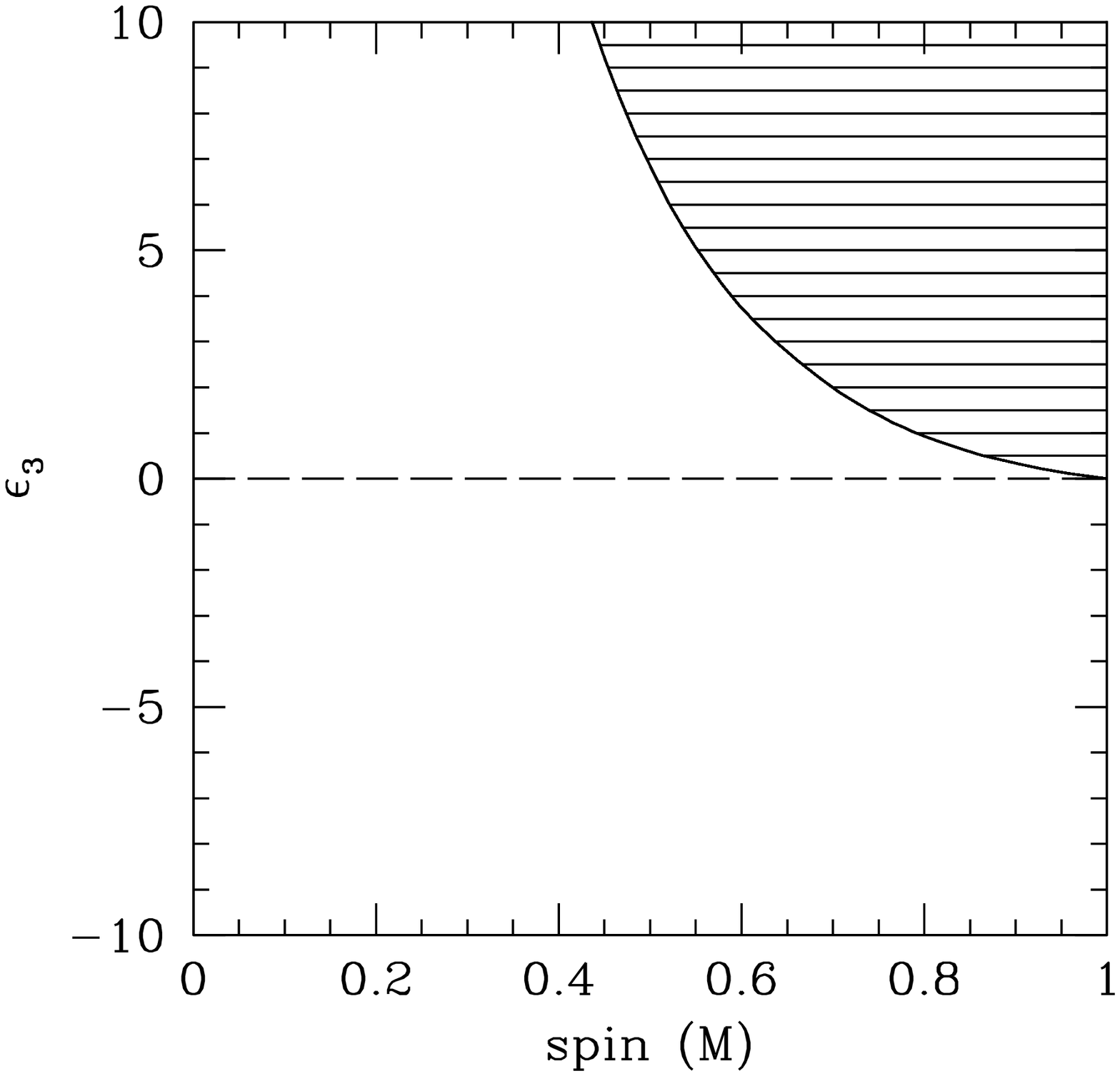,height=2.7in}
\psfig{figure=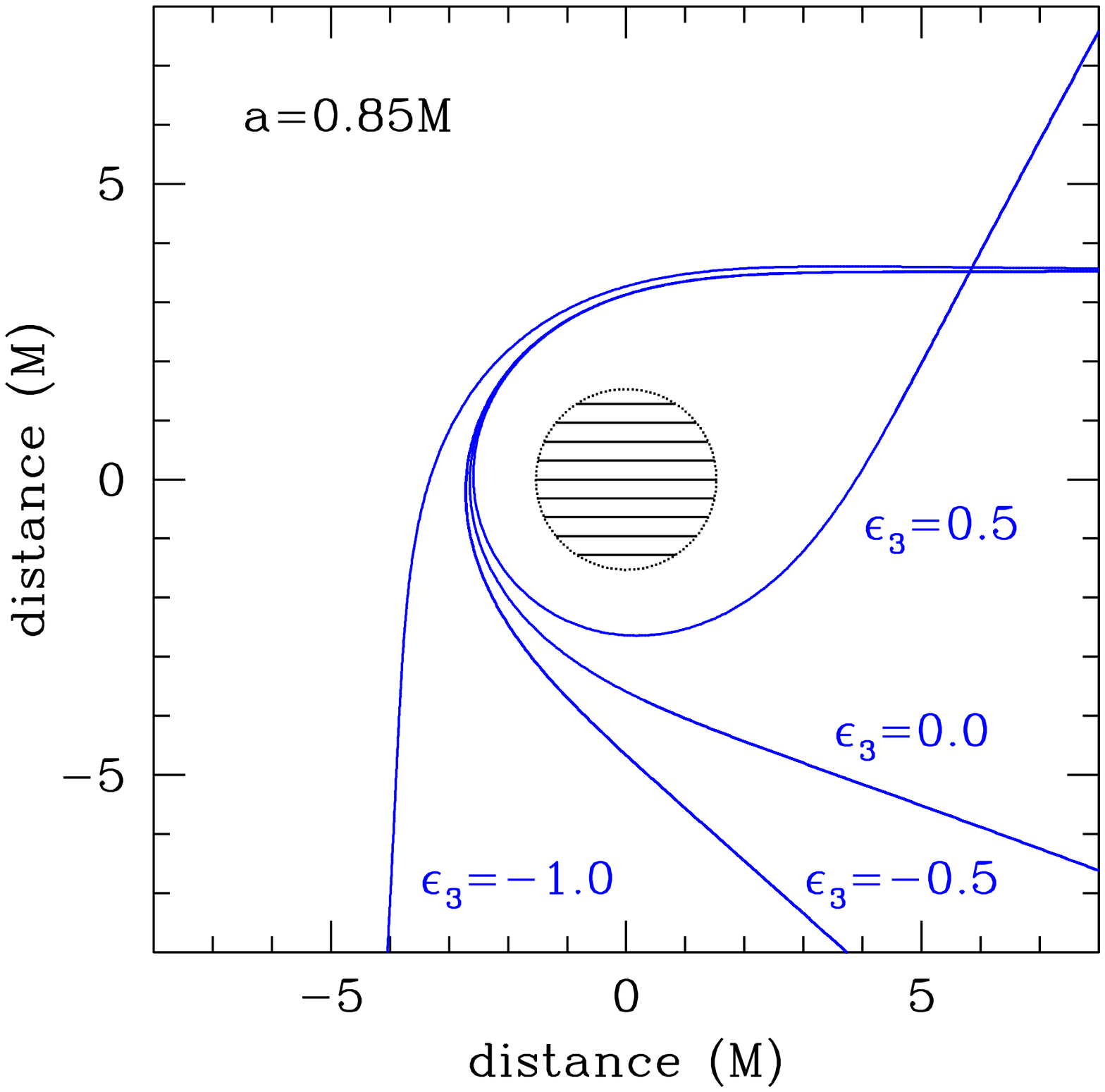,height=2.7in}
\end{center}
\caption{Left: Values of the parameter $\epsilon_3$ versus the spin $a$, for which the central object is a black hole. The shaded region marks the excluded part of the parameter space where this object is a naked singularity. The dashed line corresponds to a Kerr black hole \cite{JP11}. Right: Trajectories of photons lensed by a black hole with a (counterclockwise) spin $a=0.85M$ for several values of the parameter $\epsilon_3$. The shaded region corresponds to the event horizon of a Kerr black hole of equal spin.}
\label{f:maxdat}
\end{figure}

In \cite{JP11}, we analyzed several of the key properties of our black hole metric as a function of the spin $a$ and of one additional free parameter $\epsilon_3$. The left panel in Figure~\ref{f:maxdat} shows the range of the spin and the parameter $\epsilon_3$, for which our metric describes a black hole. The shaded region marks the part of the parameter space where the event horizon is no longer closed, and the black hole becomes a naked singularity. The right panel of Figure~\ref{f:maxdat} shows the gravitational lensing experienced by photons on an orbit in the equatorial plane that approach the black hole closely for several values of the deviation parameter $\epsilon_3$.

\begin{figure}[ht]
\begin{center}
\psfig{figure=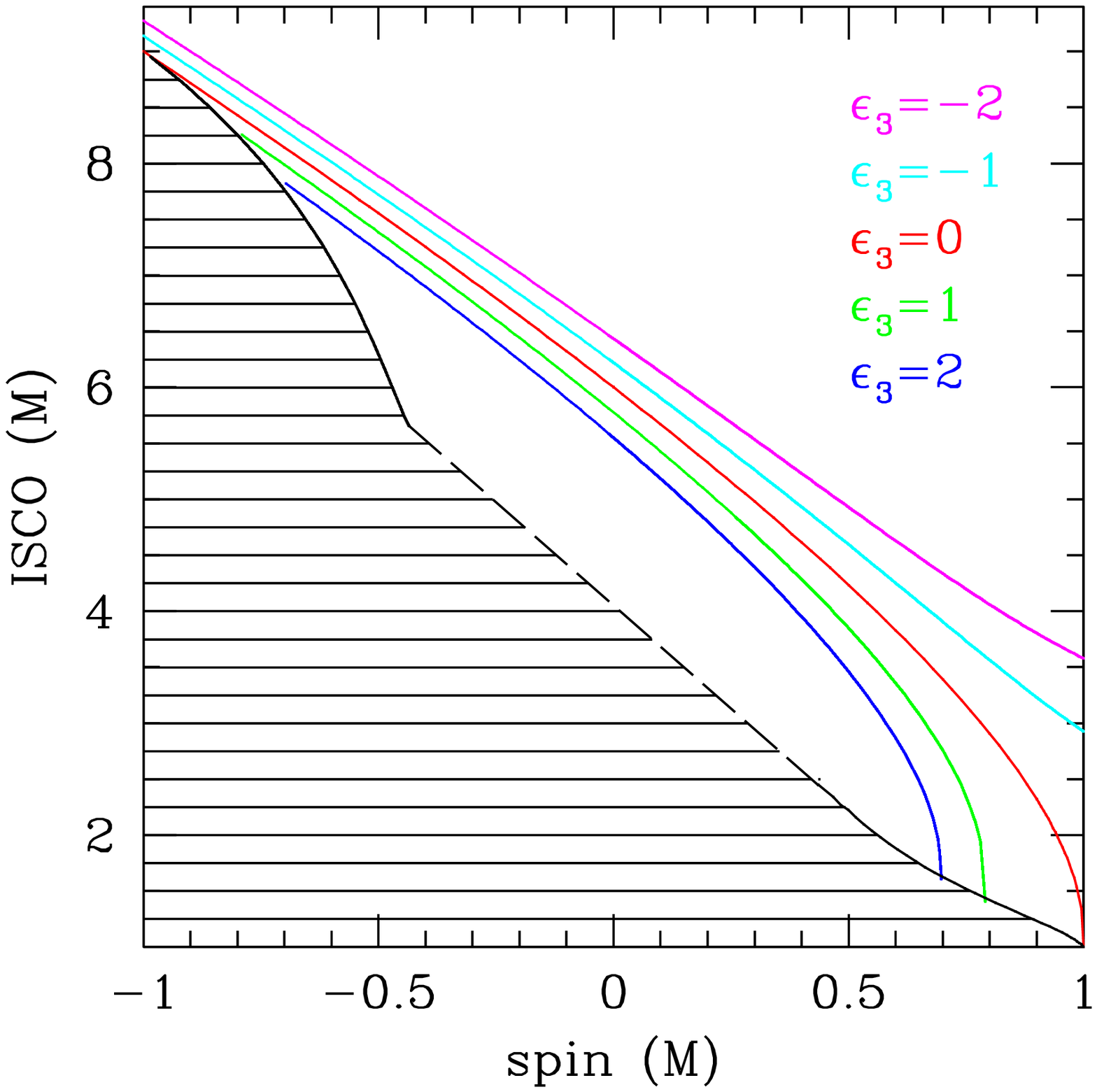,height=2.7in}
\psfig{figure=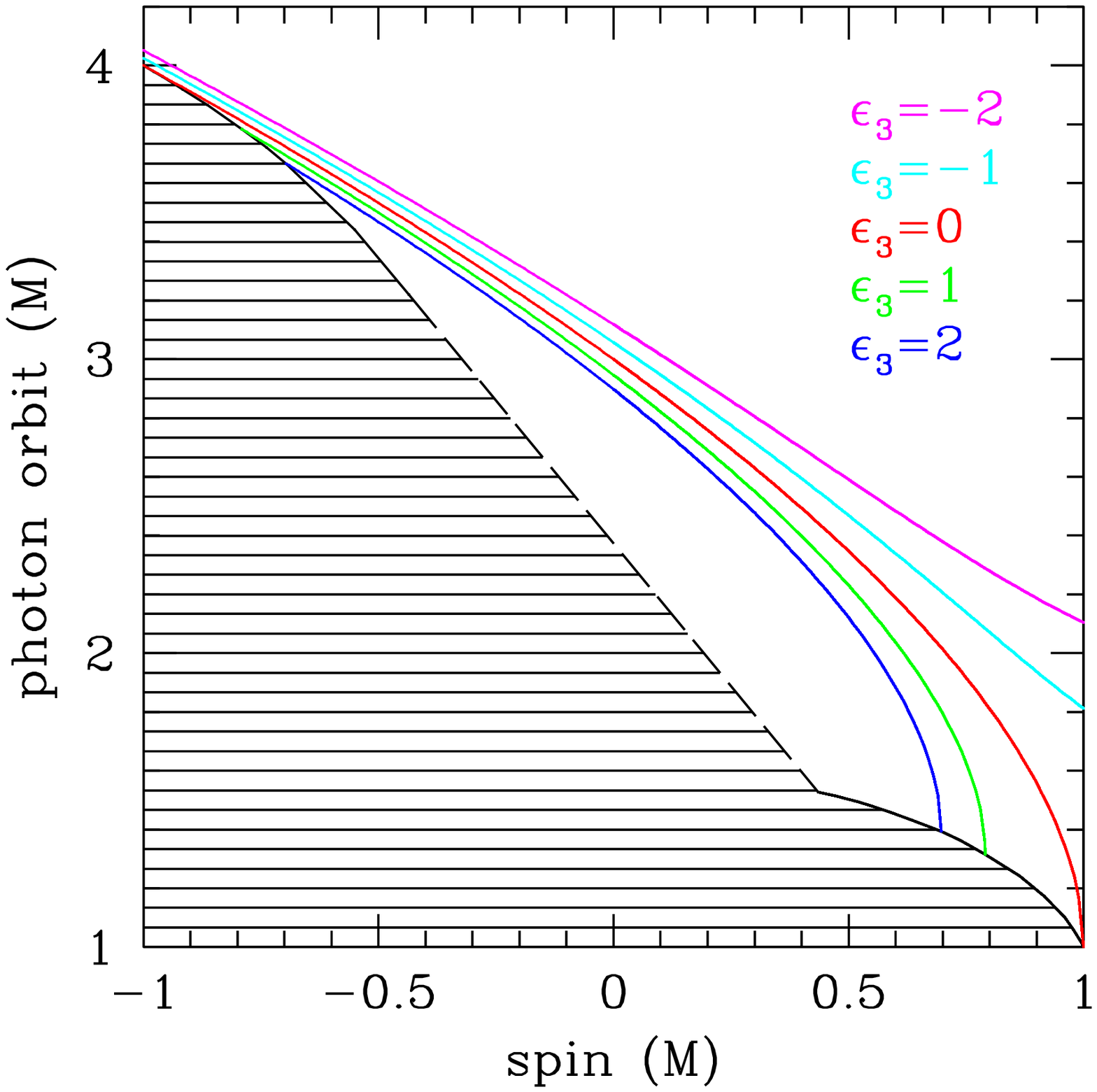,height=2.7in}
\end{center}
\caption{Radius of (left) the ISCO and of (right) the circular photon orbit as a function of the spin $a$ for several values of the parameter $\epsilon_3$. The radius of the ISCO and the circular photon orbit decrease with increasing values of the parameter $\epsilon_3$. The shaded region marks the excluded part of the parameter space \cite{JP11}.} 
\label{f:isco}
\end{figure}

\begin{figure}[ht]
\begin{center}
\psfig{figure=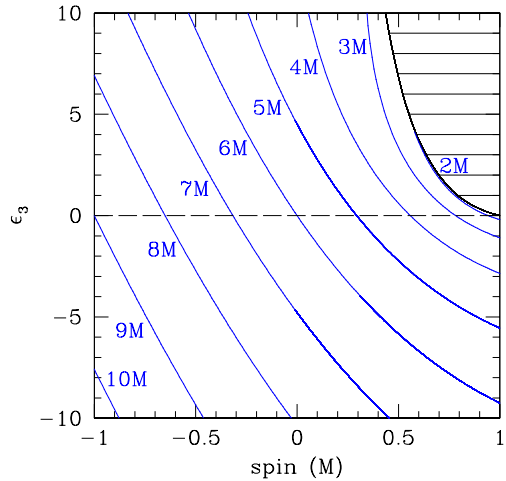,height=2.9in}
\end{center}
\caption{Contours of constant radius of the ISCO for values of the spin $-1\leq a/M \leq1$ and of the parameter $-10\leq \epsilon_3 \leq 10$. The radius of the ISCO decreases for increasing values of the spin and the parameter $\epsilon_3$. The shaded region marks the excluded part of the parameter space. The dashed line corresponds to the parameter space for a Kerr black hole \cite{JP11}.}
\label{f:iscocontours}
\end{figure}

In Figure~\ref{f:isco}, we plot the radius of the ISCO and of the circular photon orbit, respectively, as a function of the spin for several values of the parameter $\epsilon_3$. The location of both orbits decreases with increasing values of the spin and of the parameter $\epsilon_3$. In Figure~\ref{f:iscocontours}, we plot contours of constant ISCO radius as a function of the spin and the parameter $\epsilon_3$. The location of these orbits depends significantly on the value of the deviation parameter $\epsilon_3$.

\section{Testing the No-Hair Theorem with VLBI Imaging of Sgr~A$^*$}

In \cite{PaperII}, we explored in detail the effects of a violation of the no-hair theorem for VLBI imaging using a quasi-Kerr metric \cite{GB06}. This metric can be used to accurately describe Kerr-like black holes up to a spin of about $a\leq0.4M$. For Sgr~A$^*$, this spin range might already be sufficient ($a\leq0.3M$; \cite{Broderick09,Broderick10}).

The location of the circular photon orbit determines the size of the shadow of Sgr~A$^*$ (see \cite{Bardeen73,Falcke00}). VLBI observations are expected to image the shadow of Sgr~A$^*$ and to measure the mass, spin, and inclination of this black hole (e.g., [40-42]
). In addition to these parameters, the shape of the shadow also depends uniquely on the value of the deviation parameter \cite{PaperII}. In practice, however, these measurements will be model dependent (e.g., \cite{BL09}) and affected by finite telescope resolution (e.g., \cite{Falcke00,Takahashi04}). Therefore, VLBI imaging may have to be complemented by additional observations such as a multiwavelength study of polarization (\cite{BL06}; see also \cite{SK}).

In an optically thin accretion flow such as the one around Sgr~A$^*$ at sub-mm wavelengths (e.g., \cite{Broderick09}), photons can orbit around the black hole several times before they are detected by a distant observer. This produces an image of a ring that can be significantly brighter than the underlying flow thanks to the long optical path of the contributing photons (e.g., \cite{BD05}). In \cite{PaperII}, we showed that the shape and location of this ``ring of light'' depends directly on the mass, spin, inclination, and the deviation parameter of the black hole (see Figure~4). 

\begin{figure*}[ht]
\begin{center}
\psfig{figure=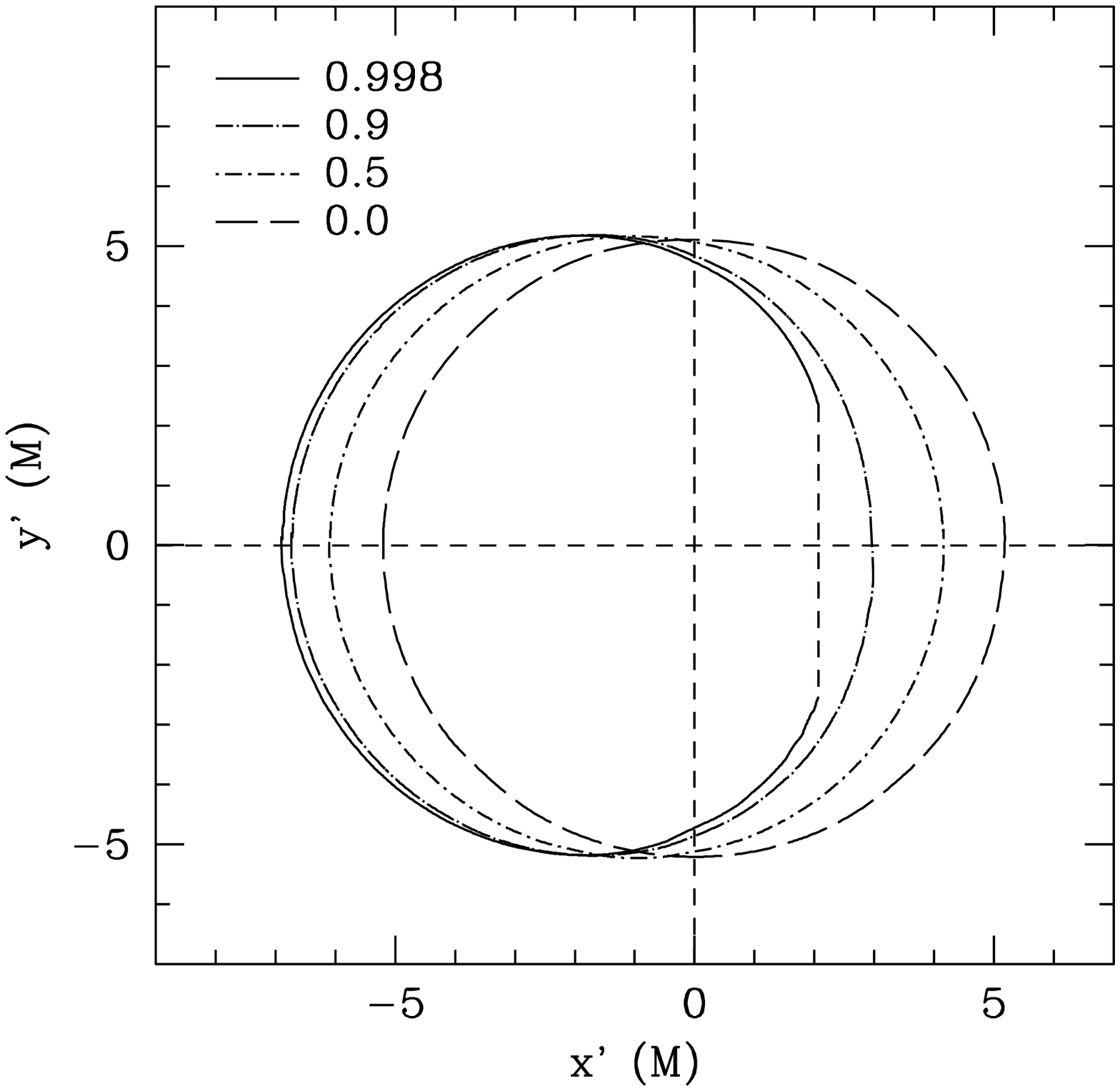,height=2.7in}
\psfig{figure=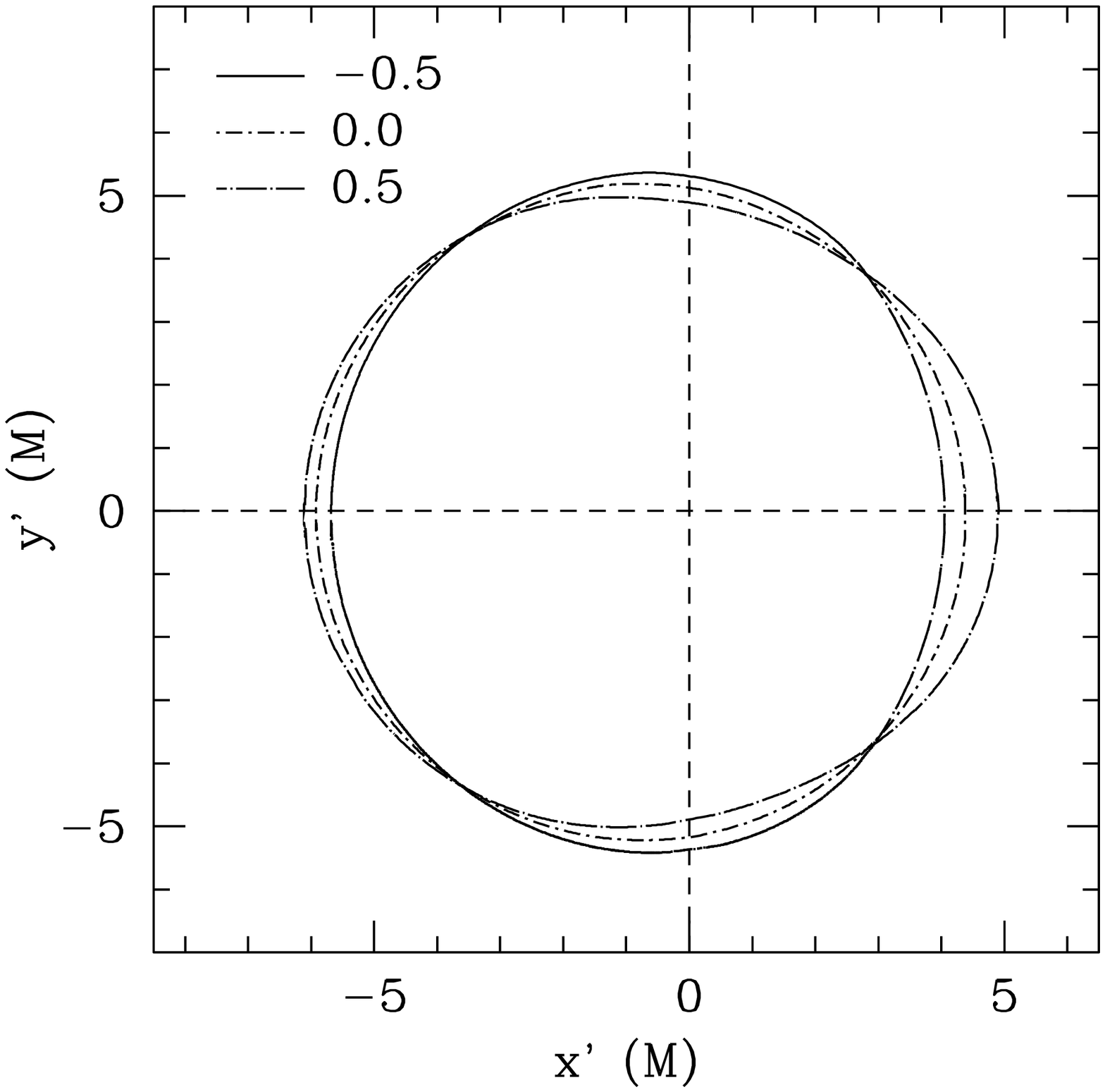,height=2.7in}
\end{center}
\caption{Images of rings of light of (left) a Kerr and (right) a quasi-Kerr black hole at an inclination $\cos i=0.25$. Increasing values of the spin cause a displacement of the ring in the image plane, but the ring remains (nearly) circular for values of the spin $a\lesssim0.9M$ (left panel). Nonzero values of the deviation parameter $\epsilon$ (right panel) lead to an asymmetric ring image \cite{PaperII}.}
\label{}
\end{figure*}

 The diameter of the ring of light as observed by a distant observer depends predominantly on the mass of the black hole and is nearly constant for all values of the spin and disk inclination as well as for small values of the deviation parameter. For nonzero values of the spin of the black hole, the ring is displaced off center in the image plane. In all cases, the ring of a Kerr black hole remains nearly circular except for very large values of the spin $a\gtrsim0.9M$. However, if Sgr~A$^*$ is not a Kerr black hole, the ring becomes asymmetric in the image plane. This asymmetry is a direct measure for a violation of the no-hair theorem (see Figure~5).

\begin{figure*}[ht]
\begin{center}
\psfig{figure=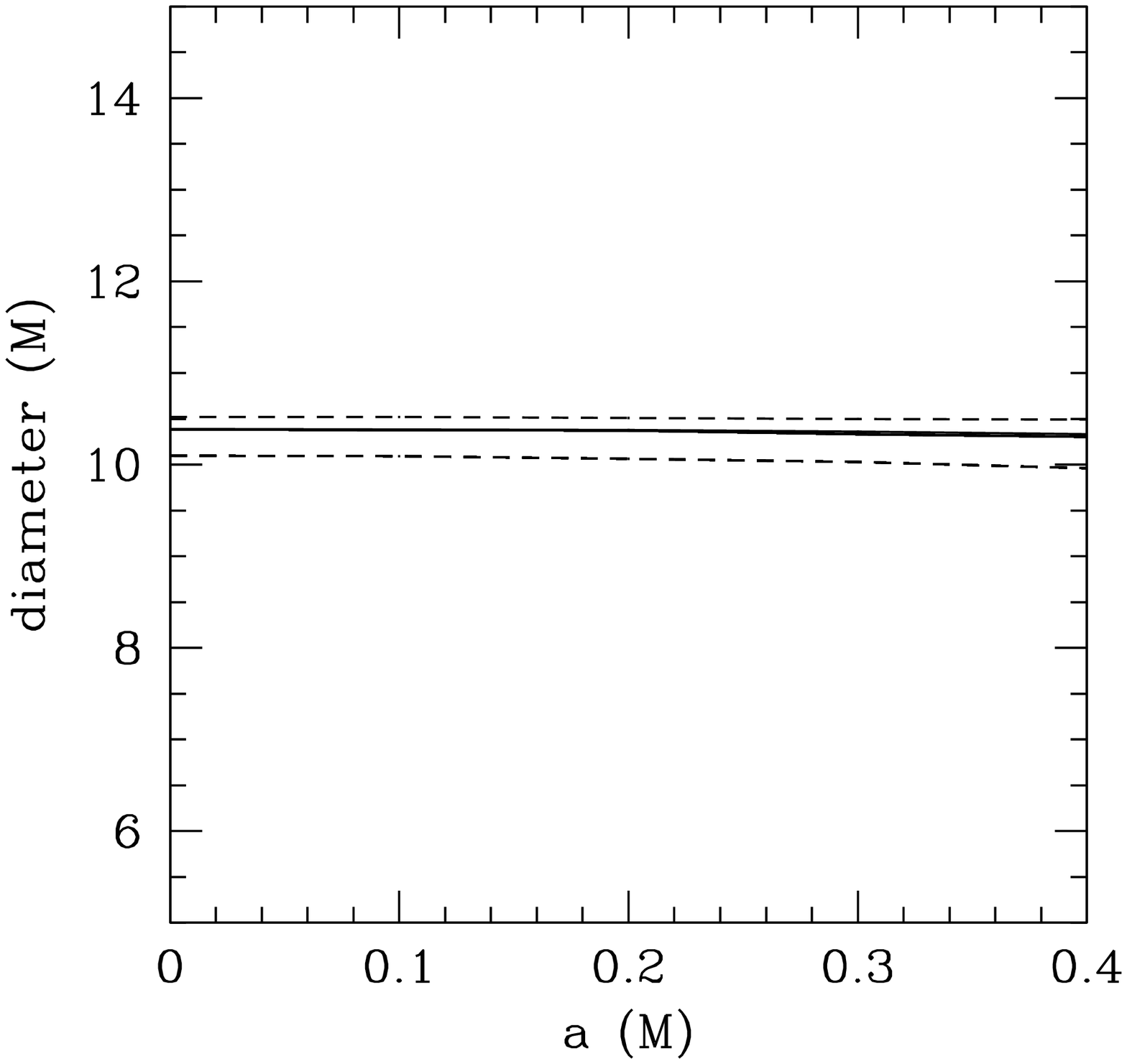,height=2in}
\psfig{figure=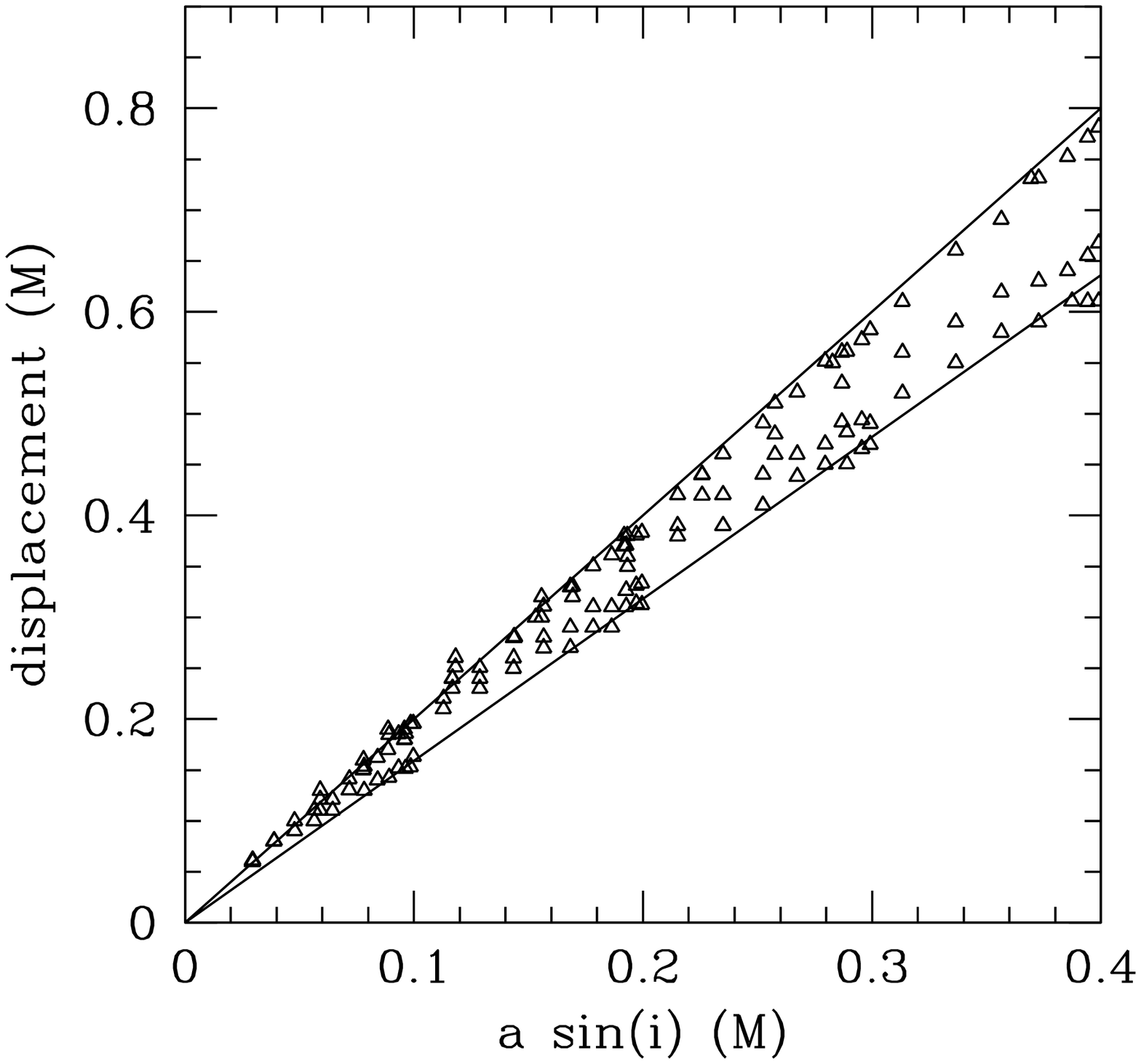,height=2in}
\psfig{figure=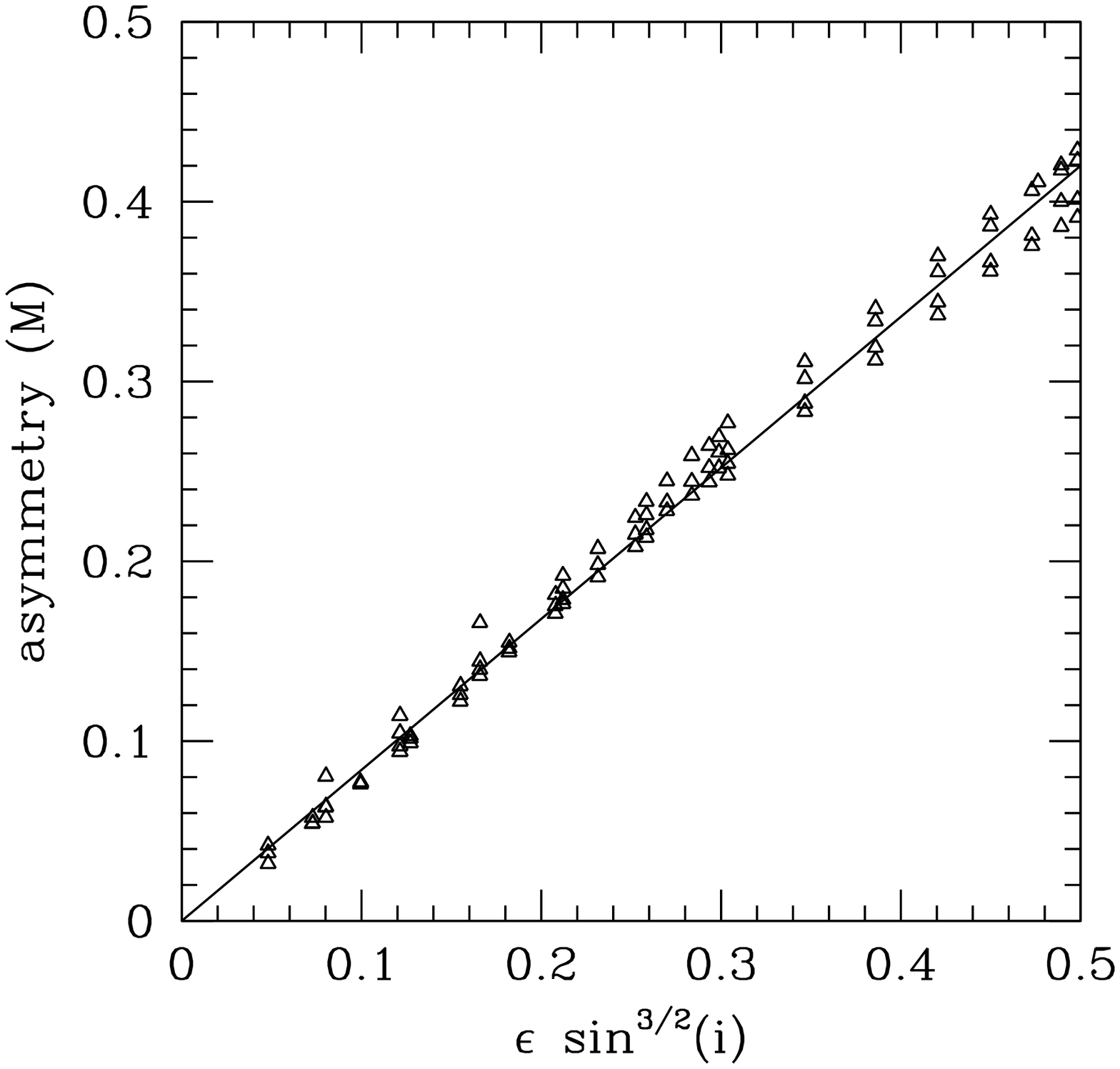,height=2in}
\end{center}
\caption{Left: The ring diameter versus spin for inclinations $17^\circ\leq i \leq86^\circ$ for a Kerr black hole (solid lines) and for a quasi-Kerr black hole with a value of the deviation parameter $\epsilon=0.5$ (dashed lines). The diameter is practically independent of the spin, inclination, and deviation parameter with a constant value of $\simeq$10.4$M$ for a Kerr black hole. Center: The displacement of the ring of light as a function of $a\sin i$ for various values of the parameter $0\leq\epsilon\leq0.5$. The displacement depends only weakly on the parameter $\epsilon$. Right: The ring asymmetry versus $\epsilon\sin^{3/2} i$ for various inclinations $17^\circ\leq i \leq86^\circ$ and $0.0\leq a/M \leq0.4$. The asymmetry is nearly independent of the spin and hence provides a direct measure of a violation of the no-hair theorem \cite{PaperII}.}
\label{}
\end{figure*}

\section{Combining Strong-Field with Weak-Field Tests of the No-Hair Theorem}

\begin{figure}[ht]
\begin{center}
\psfig{figure=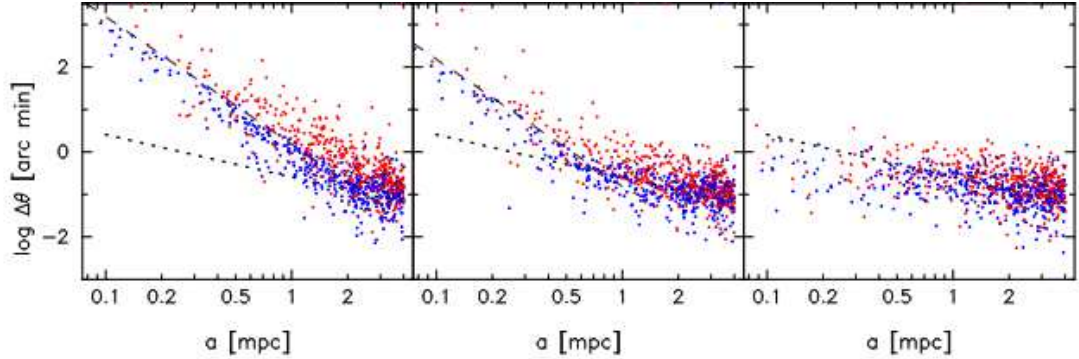,width=6in}
\end{center}
\caption{Evolution of the orbital angular momenta of stars around Sgr~A$^*$ due to frame-dragging (dashed lines) and stellar perturbations (dotted lines) as measured by the angle $\Delta\theta$ as a function of the orbital semi-major axis $a$. The three panels correspond to a Kerr black hole with a spin (left) $a=M$, (middle) $a=0.1M$, and (right) $a=0$ \cite{Merritt10}.}
\label{f:Merritt}
\end{figure}

In addition to a strong-field test of the no-hair theorem with VLBI imaging of Sgr~A$^*$, there exist two other promising possibilities for performing such a test in the weak-field regime. The presence of a nonzero spin and quadrupole moment independently leads to a precession of the orbit of stars around Sgr~A$^*$ at two different frequencies, which can be studied with parameterized post-Newtonian dynamics \cite{Will08,Will93}. Merritt et al. \cite{Merritt10} showed that the effect of the quadrupole moment on the orbit of such stars is masked by the effect of the spin for the group of stars known to orbit Sgr~A$^*$. However, if a star can be detected within $\sim1000$ Schwarzschild radii of Sgr~A$^*$ and if it can be monitored over a sufficiently long period of time, this technique may also measure the spin (see Fig.~\ref{f:Merritt}; \cite{Merritt10}) and even the quadrupole moment \cite{Merritt10} together with the already obtained mass \cite{Sgr}. Future instruments, such as GRAVITY \cite{Bartko09}, may be able to resolve the orbits of such stars providing an independent test of the no-hair theorem.

\begin{figure}[ht]
\begin{center}
\psfig{figure=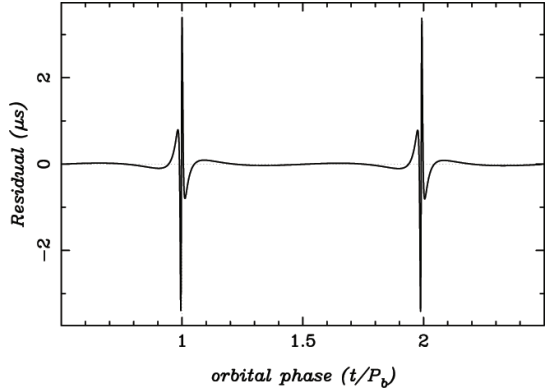,width=2.8in}
\end{center}
\caption{Typical timing residuals for a radio pulsar in an orbit around a black hole with a mass of $10^4 M_\odot$ due to the presence of a nonvanishing quadrupole moment \cite{WK99}. For Sgr~A$^*$, such residuals will have a similar shape but with a larger amplitude.}
\label{f:pulsar}
\end{figure}

Yet another weak-field test can be performed by the observation of a radio pulsar on an orbit around Sgr~A$^*$. If present, timing observations may resolve characteristic spin-orbit residuals that are induced by the quadrupole moment and infer its magnitude (see Fig.~\ref{f:pulsar}; \cite{WK99}). Recent surveys set an upper limit for the existence of up to 90 pulsars within the central parsec of the galaxy \cite{Macquart10} making this technique a promising third approach for testing the no-hair theorem with Sgr~A$^*$.

The fundamental properties of the black hole in the center of our galaxy can be probed with three different techniques. The combination of the results of all three approaches will lead to a secure test of the no-hair theorem with Sgr~A$^*$.

\end{document}